%% file: main.tex
\documentclass[journal]{IEEEtran} 
\usepackage{cite}
\usepackage{graphicx} 
\usepackage{pifont} \usepackage{amsmath,amssymb,amsthm} 
\usepackage{makecell}
\usepackage{multirow}
\usepackage{mathtools} \usepackage{algorithmic} \usepackage{algorithm} \usepackage{caption} \usepackage{subcaption} \usepackage{epstopdf} \usepackage{enumitem} \usepackage{tabularx} \usepackage{lipsum} \usepackage{hyperref} \usepackage{svg} \usepackage{float}  \usepackage{graphicx,float} \usepackage{pgfplots} \usepackage{pgfplotstable} \usepackage{subcaption} \usepackage{adjustbox} \pgfplotsset{compat=1.18} \usepackage{multirow} % For multirows
\usepackage{graphicx} % For the rotati
\usepackage{tabularx,booktabs} 
\usepackage{pifont} \usepackage{relsize} \usepackage[utf8]{inputenc}
\usepackage{diagbox} \usepackage{pdflscape} \usepackage{longtable} \usepackage{array} \usepackage{geometry}  \usepackage[table]{xcolor} \usepackage{tikz} \usepackage{booktabs} \usepackage{listings} \usepackage{pifont} \usepackage{ifthen}  \usepackage{xcolor} \usepackage[table]{xcolor} \usepackage{longtable} \usepackage{geometry} \usepackage{array} \usepackage{caption} \usepackage{booktabs} \usepackage{arydshln} \usepackage{enumitem} \usepackage{relsize} \renewcommand{\tablename}{Tab.} \usepackage[skip=5pt]{caption} \newcolumntype{C}{>{\centering\arraybackslash}X}    \newcommand{\nuc}{\newcommand} \theoremstyle{plain}    \theoremstyle{definition}   \theoremstyle{remark}  \usepackage{verbatim} \usepackage{tablefootnote} 
\usepackage[font=footnotesize]{caption}
\begin{document} 
\title{Data and Context Matter: Towards Generalizing AI-based Software Vulnerability Detection}

\author{Rijha~Safdar, Danyail~Mateen, Syed~Taha~Ali, Umer~Ashfaq and Wajahat~Hussain% <-this % stops a space
\thanks{R. Safdar, S.T. Ali and W. Hussain are with School of Electrical Engineering and Computer Science, National University of Sciences and Technology, Islamabad, Pakistan, 44000. e-mail: rsafdar.dphd19seecs@seecs.edu.pk , e-mail: taha.ali@seecs.edu.pk, email:wajahat.hussain@seecs.edu.pk} }
 
\maketitle 
% We present DeepVul, the most diverse publicly available C/C++ vulnerability dataset to date, containing 236K function-level samples across 92 CWE types. Our automated pipeline updates prior datasets (e.g., Big-Vul, last updated in 2019) with CVEs up to May 2025 and removes mislabeled samples (14% in Big-Vul, 4.6% in Devign, 4.1% in REVEAL). We further introduce hard negative samples—near-identical secure/vulnerable code pairs—to improve model generalization. Benchmarking five LLMs (125M–7B parameters) on Big-Vul, DeepVul, and three unseen datasets shows consistent gains in recall and robustness, with UniXcoder variants performing best overall. 

\begin{abstract} AI-based solutions demonstrate remarkable results in identifying vulnerabilities in software, but research has consistently found that this performance does not generalize to unseen codebases. In this paper, we specifically investigate the impact of model architecture, parameter configuration, and quality of training data on the ability of these systems to generalize.

For this purpose, we introduce VulGate, a high quality state of the art dataset that mitigates the shortcomings of prior datasets, by removing mislabeled and duplicate samples, updating new vulnerabilities, incorporating additional metadata, integrating hard samples, and including dedicated test sets. We undertake a series of experiments to demonstrate that improved dataset diversity and quality substantially enhances vulnerability detection. We also introduce and benchmark multiple encoder-only and decoder-only models. We find that encoder-based models outperform other models in terms of accuracy and generalization. Our model achieves \textbf{6.8\%} improvement in recall on the benchmark BigVul dataset and outperforms others on unseen projects, demonstrating enhanced generalizability. Our results highlight the role of data quality and model selection in the development of robust vulnerability detection systems. Our findings suggest a direction for future systems with high cross-project effectiveness.

%data quality and model architecture on the generalizability of vulnerability detection systems. By generalization we mean ability of high vulnerability detection performance across different C/C++ software projects not seen during training. Through a series of experiments, we demonstrate that greater dataset diversity and quality substantially enhance detection performance. Additionally, we compare multiple encoder-only and decoder-only models, finding that encoder based models outperform in terms of accuracy and generalization. Our model achieves \textbf{6.8\%} improvement in recall on the benchmark BigVul\cite{1} dataset and also outperforming on unseen projects, demonstrating enhanced generalizability. These results highlight the role of data quality and model selection in the development of robust vulnerability detection systems. Our findings suggest a direction for future systems with high cross-project effectiveness. 
\end{abstract} 


\IEEEpeerreviewmaketitle 
% \input acronyms.tex
\input macros.tex 
\input 1-Intro.tex
\input 2-a_BG.tex 
\input 2-b_LR.tex

\input 3-DataandModel.tex 
\input 4-Exp.tex

\input 5-discussion.tex
\input 6-end.tex
\bibliographystyle{IEEEtran} \bibliography{IEEEabrv,ref} 
\end{document}

%% file: macros.tex
% ABCD
%%%%%%%%%%%%
\nuc{\bA}{\bar{A}}
\nuc{\mb}{\mathcal{B}}
\nuc{\mc}{\mathcal{C}}
\nuc{\cc}{\cite}
\nuc{\cS}{\mathcal{S}}
\nuc{\Del}{\Delta}
\nuc{\del}{\delta}

% EFGH
%%%%%%%%%%%%
\nuc{\ben}{\begin{enumerate}}
\nuc{\een}{\end{enumerate}}
\nuc{\EB}{\begin{eqnarray*}}
\nuc{\EBN}{\begin{eqnarray}}
\nuc{\EE}{\end{eqnarray*}}
\nuc{\EEN}{\end{eqnarray}}
\nuc{\EQ}{\begin{equation*}}
\nuc{\EQN}{\begin{equation}}
\nuc{\EN}{\end{equation*}}
\nuc{\ENN}{\end{equation}}
\nuc{\E}{\mathbf{E}}
\nuc{\eD}{e^{\Del}}
\nuc{\eps}{\epsilon}
\nuc{\Gam}{\Gamma}
\nuc{\gam}{\gamma}
\nuc{\bF}{\bar{F}}
\nuc{\half}{\frac{1}{2}}

% IJKL
%%%%%%%%%%%%
\nuc{\kk}{{k|k}}
\nuc{\la}{\leftarrow}
\nuc{\lab}{\label}
\nuc{\lam}{\lambda}

% MNOP
%%%%%%%%%%%%
\nuc{\um}{^{-1}}
\nuc{\nn}{\nonumber}

% QRST
%%%%%%%%%%%%
\nuc{\Ra}{\Rightarrow}
\nuc{\ra}{\rightarrow}
\nuc{\Sg}{\Sigma}
\nuc{\sg}{\sigma}
\nuc{\tbf}{\textbf}
\nuc{\tA}{\tilde{A}}
\nuc{\tB}{\tilde{B}}
\nuc{\tC}{\tilde{C}}
\nuc{\te}{\tilde{e}}

% UVWXYZ
%%%%%%%%%%%%
\nuc{\ul}{\underline}
\nuc{\bu}{\bar{u}}
\nuc{\uD}{u^{\Del}}
\nuc{\tu}{\tilde{u}}
\nuc{\mv}{\mathcal{V}}
\nuc{\bx}{\bar{x}}
\nuc{\hx}{\hat{x}}
\nuc{\hxD}{\hx^{\Del}}
\nuc{\tx}{\tilde{x}}
\nuc{\xa}{x^a}
\nuc{\xD}{x^{\Del}}
\nuc{\ya}{y^a}
\nuc{\yD}{y^{\Del}}
\nuc{\by}{\bar{y}}
\nuc{\ty}{\tilde{y}}
\nuc{\zD}{z^{\Del}}

%% file: 1-Intro.tex
\section{Introduction}
% motivation #1: vulnerabilities volume-out-of-control
With the rapid growth in digitization and software applications and systems in recent years, the issue of software vulnerabilities has become a critical concern. In 2024, a record-breaking 40,000 Common Vulnerabilities and Exposures (CVEs) were published—an average of ~108 per day—marking a 38\% increase over 2023 (with 28,818 CVEs) \cite{Baran2025Jan}. This number is already dramatically increasing: the first half of 2025 has witnessed an average of 131 CVEs per day \cite{BibEntry2025Jul}. In the open-source software ecosystem, which underpins a wide range of industries, including finance, energy, aerospace, and healthcare, a recent study found a surge of 98\% per year in reported vulnerabilities \cite{akhavani2025opensourceopenthreats}.

% motivation #2: real-world impact
The real-world impact is also significant. Whereas critical vulnerabilities grew 37\% in 2024, the known exploited vulnerabilities jumped 96\% \cite{McDade2025Jul}. Furthermore, 23.6\% of exploited vulnerabilities were leveraged on or before public disclosure, and half were typically exploited within 192 days \cite{BibEntry2025Aug}. Meanwhile, the average time to fix software vulnerabilities has grown to eight and a half months, an increase of 47\% over the last five years \cite{Coker2025Aug}.

% automated detection and LLMs
These alarming trends highlight the need for reliable automated solutions to identify software vulnerabilities at scale. The emergence of large language models (LLMs) appears to be a promising development in this direction. Although originally trained on a massive corpora of natural language, these tools demonstrate strong performance in generating code, identifying bugs, and patching vulnerabilities \cite{26, 28, 29}. Popular coding tools, including GitHub Copilot \cite{31} and Cursor \cite{32}, have integrated LLMs for real-time debugging. 

% where we are with research
The research community is actively investigating the use of LLMs, particularly pre-trained models, to identify logic bugs and subtle security vulnerabilities \cite{27} \cite{30}. Whereas existing architectures demonstrate remarkable results on training and testing datasets, a fundamental limitation is that this \textbf{performance does not generalize} to unseen codebases \cite{33}. For instance, state of the art systems VulDeePecker \cite{3} and SySeVR \cite{8} report F1 scores of 85.4\% and 84.4\% respectively on benchmark datasets, but encounter severe performance degradation on the more realistic \textsc{Re}\textlarger{V}\textsc{eal} dataset, scoring 12.18\% and 19.05\% respectively \cite{9}. This lack of generalization limits practical applicability of LLMs to real-world scenarios where software code can vary considerably in terms of structure, semantics, and vulnerabilities. 

% focus of our study: reason #1
Whereas prior work has speculated on multiple reasons for this lack of generalization, in this paper we undertake a rigorous empirical study to evaluate the role of three key factors: first, multiple researchers have indicated that \textbf{datasets used in prior assessments are problematic}, featuring significant label noise, duplicated samples, and class bias \cite{9}, \cite{38}, \cite{52}. This results in overly optimistic assessment outcomes, with models that seem accurate in benchmarks yet perform poorly in practice.

Second, \textbf{most datasets lack sufficient diversity}, with a bias toward limited types of projects or vulnerability types \cite{38}, \cite{53}. This narrow scope prevents model’s ability to learn generalizable patterns, leading to missed or misclassified vulnerabilities outside the dataset’s scope.

The third factor is the \textbf{choice of models and their configuration}. Our experiments demonstrate that newer transformer architectures, pre-trained on code, dramatically outperform GNN and deep-learning solutions. Parameter configuration is also critical: prior models had limited context window size, mostly restricted to 512 length, which constraints their ability to capture long range dependencies in code. A small context window forces the model to truncate input, omitting critical code pertaining to a vulnerability \cite{43}.

% our exact contributions
Our precise contributions in this paper are:

\begin{enumerate}
% [which datasets does this comprise? %ans
% What are modules/cpmponents?
% How much more data/CVEs were added when bringing it up to date?
% How many corrections were made?
% How many man hours did cleaning exercise take? Mislabeling, deduplication, and balancing?
% 92 CWE types as compared to how many in others?
% Can we introduce GenVul as part of this? How does GenVul enable "cross project generalization? What is the size of this? 
% How long did manual verification take?
% What are the other external applications for VulGate? novel element]
 % $\sim$
\item We introduce \textbf{VulGate}, a large scale, rigorously curated dataset to train models for software vulnerability detection, classification and patching. VulGate unifies and cleans samples from established corpora (including, Devign, BigVul, \textsc{Re}\textlarger{V}\textsc{eal}, VDISC, D2A, CVEfixes, CrossVul, DiverseVul, PrimeVul and MegaVul). The dataset contains 236,663 function-level samples with newly-scraped samples up to May, 2025, along with cleaned vulnerability-related diff-based annotations. Compared to the widely used BigVul dataset, VulGate expands coverage from 91 to 180 CWE types, adds 48,027 new samples (25.5\% increase), and incorporates new subsets manually verified by security experts. It also includes a 500-sample expert-verified test set \textbf{VulGate$^{+}$} explicitly designed to evaluate cross-project generalization.

% which train the model to recognize subtle distinctions between secure and vulnerable patterns.

\item We deliberately incorporate \textbf{hard negative samples} in VulGate—function pairs with high semantic similarity ($\geq 0.90$ cosine similarity, nearly 90\% in training set). These pairs typically exhibit minimal syntactic difference, often just one line or token (e.g. replacing \textit{strcpy} with \textit{strncpy}), reflecting real-world CVEs where vulnerabilities are caused with minimal changes. Hard negatives force models to learn semantic patterns as opposed to superficial syntax, thereby improving generalization to unseen codebases. Prior work shows that incorporating hard negatives consistently boosts performance (e.g., Li et al.\cite{67} reported +0.046 MRR improvement for CodeBERT on the Java subset of CodeSearchNet \cite{58,61}). This is a novel contribution in the context of vulnerability detection, which has not been explored in prior work.

\item To assess model architecture and the particular role of context window size in vulnerability detection, we undertake \textbf{comprehensive benchmarking} exercises for five state of the art LLMs. These comprise encoder-only models (CodeBERT, UniXcoder variants) and decoder-only models (CodeGPT-2, Code Llama), spanning the range of lightweight to large-scale deployments with parameter counts from 125M–7B. We find that extended context window along with encoder-only architecture type jointly influence performance as their masked language modeling training allows the model to process whole code sequences simultaneously.

\end{enumerate}

% results overview - para 1
% significant improvement on vulnerability detection []

% Our results show that VulGate, combined with targeted fine-tuning [you mean the context window?] improves both vulnerability detection performance and generalization. On BigVul, UniXcoder-Base-Nine [introduce/justify?] achieved a 6.8\% increase in recall over CodeBERT, with a 94.73\% F1. [How/why is this a big deal?]

Our results indicate that encoder-only models capable of handling longer contexts (e.g., UniXcoder-Base-Nine with 1024-token window) and trained on VulGate, demonstrate significantly improved performance. On the widely-used BigVul benchmark, UniXcoder-Base-Nine achieves an F1 score of 94.73\% and 6.8\% improvement in absolute recall over CodeBERT. UniXcoder-Base-Nine also consistently outperforms baselines, including a 47 point F1 improvement on a synthetic code subset.

Moreover, our results represent a breakthrough in terms of generalization. Recent benchmark studies demonstrate precipitous performance loss of ~40-70\% for state of the art models transitioning from seen to unseen datasets \cite{38} \cite{33}. In contrast, our performance loss is an order of magnitude lower, dropping only 4–6\% on the PrimeVul benchmark dataset. To the best of our knowledge, this represents the most robust cross-dataset generalization results reported in the literature to date.

The remainder of the paper is organized as follows: Section II includes background material on vulnerability detection. In Section III we undertake a literature review of leading datasets. Section IV introduces our dataset VulGate and describes the data construction pipeline. In Section V, we review vulnerability detection models in the research literature and describe our benchmarking and fine-tuning experiments. In Section VI we discusses insights. Section VII concludes the paper with a summary of findings and future work.

%% file: 2-a_BG.tex
\section{Background}

Here we provide an overview of automated techniques for software vulnerability detection.

%The increasing dependence on open-source and sophisticated software development have increased the demand for automated tools that can detect software vulnerabilities, particularly in C/C++ applications. These powerful languages are prone to memory/low-level flaws that can lead to serious security vulnerabilities.

\subsection{Traditional Vulnerability Detection Solutions}
%Traditional Vulnerability Detection Techniques
Conventional vulnerability detection solutions comprise static and dynamic analysis, hybrid solutions and symbolic execution techniques.

% static analysis
\textbf{Static analysis} tools (such as Fortify\cite{19}, Clang\cite{20}, Cppcheck\cite{21} etc.) review source code, analyzing program structure, data and control flow, syntax accuracy, etc. and identify well-known bugs and vulnerabilities \cite{18}. The code itself is not executed in a live environment, which allows more subtle flaws to escape detection. Static analyzers also have limited contextual understanding of the code and often suffer from high false positive (FP) rates.

Industry guidelines recommend keeping FPs below 20–30\% to make a tool acceptable to users. An analysis of popular open-source embedded software using Github's CodeQL reported an FP rate of around 23\% \cite{shen2023empirical}. Google's development environment limits FPs to 10\% based on user feedback \cite{44}. In practice, however, FP rates can be much higher: a large-scale study reports that up to 96\% of static analyzer warnings were false \cite{45}. %Such limitations results in lack of developer trust, usability and increase in computational resources.

% (e.g.? compelling numbers + sources+ prestige factor) + we observe similar trends in our own results in Table X, where FP rate is...?

% dynamic analysis
\textbf{Dynamic analysis tools} evaluate software behavior during runtime, helping identify issues like memory leaks, performance bottlenecks, and security vulnerabilities. Detection techniques include taint analysis (i.e., tracking data flow of untrusted inputs) and fuzzing (i.e., testing for robustness against abnormal inputs) \cite{24,23}. Dynamic analysis can be more effective at identifying vulnerabilities than static analysis, but this approach has various issues: it requires security experts to run applications in virtual environments. This process can be computationally expensive and time-consuming, sometimes taking days to scale across large codebases. Dynamic analysis also struggles with limitations in code coverage (only detecting issues in code paths that were executed).

% hybrid analysis
\textbf{Hybrid analysis} combines the aforementioned static and dynamic approaches by integrating run-time data extracted from dynamic analysis into a static analysis algorithm. This results in significant performance improvements (e.g., one evaluation reported 17\% improved accuracy and 25\% faster detection rate \cite{47}). However, hybrid analysis is still resource intensive, requiring run-time environment, computational resources, and oversight by security experts.

% symbolic execution
\textbf{Symbolic execution} is another traditional technique where tools (such as KLEE \cite{22}) explore program paths systematically to identify bugs. The major challenge in this regard is scalability due to path explosion: an analysis by Bessler \textit{et al.} demonstrates that the number of program paths often grows exponentially with function complexity \cite{48}. This process often requires substantial computational resources and time for constraint solving.

% motivating ML/DL
These traditional techniques, though valuable, often struggle to scale across a wide range of evolving codebases. Consequently, \textbf{machine learning} (ML) and \textbf{deep learning} (DL) approaches are gaining attention for their ability to learn vulnerability patterns, enabling more accurate, scalable, and quicker results. For instance, an ensemble learning solution, VELVET, outperforms baseline static analyzers by a factor of 4.5 and with significantly reduced FP rates \cite{49}.

\subsection{AI for Vulnerability Detection}

We refer the reader to \cite{kaniewski2025systematic} for a detailed survey on this topic. Here we provide a brief overview:

These approaches treat vulnerability detection as a classification task. ML and DL models rely on features extracted from abstract syntax trees, control flow graphs, data flow graphs, or program dependence graphs\cite{3,4}. Whereas this approach significantly improves automation, the tools required to generate input graphs are mostly language-dependent. Cross-project generalization is also an issue with models overfitting to specific datasets.

%Large Language Models and Their Limitations
\textbf{LLMS}, based on transformer architectures and pretrained on large code and natural language datasets, have also demonstrated exceptional capability to process and generate code.

However, pre-trained models require fine-tuning for vulnerability detection, the effectiveness of which in turn depends considerably on the quality and diversity of training data. As documented in our literature review (Sec.\ref{sec:literature_review}), these models give strong results on datasets on which they are trained, but performance degrades considerably on unseen codebases. As we demonstrate in this paper, a significant component of this drop in performance is due to label noise and data duplication, imbalanced datasets, and small context window.

%The models tend to perform well on the data which they are trained but when evaluated on unseen codebases, they struggle mainly due to label noise and data duplication, biases in training (imbalanced datasets) and short context window. 

%However, to use pre-trained model for the task of vulnerability detection, the model is fine-tuned. The effectiveness of the fine-tuned models greatly depend on the quality and diversity of training data. The models tend to perform well on the data which they are trained but when evaluated on unseen codebases, they struggle mainly due to label noise and data duplication, biases in training (imbalanced datasets) and short context window. These shortcomings highlight the need for both good quality datasets and architectural improvements in LLMs for reliable, generalizable vulnerability detection, which is previously not achieved in literature.

%The details of related LLM models are discussed in Section~\ref{subsec:model-selection}.
% Define checkmark and cross

%% file: 2-b_LR.tex
\section{Literature Review: Datasets}

\newcommand{\cmark}{\textcolor{green!60!black}{\ding{51}}}%
\newcommand{\xmark}{\textcolor{red}{\ding{55}}}%

% Define star ratings
\newcommand{\high}{\textcolor{blue!70!black}{\ding{72}\ding{72}\ding{72}\ding{72}\ding{72}}} % 5 stars
\newcommand{\mhigh}{\textcolor{blue!70!black}{\ding{72}\ding{72}\ding{72}\ding{72}}} % 4 stars (blue tone to distinguish)
\newcommand{\medium}{\textcolor{blue!70!black}{\ding{72}\ding{72}\ding{72}}} % 3 stars
\newcommand{\low}{\textcolor{blue!70!black}{\ding{72}\ding{72}}} % 2 star

\begin{table*}[ht]
\centering
% \scriptsize
\resizebox{\textwidth}{!}{
\begin{tabular}{|l|c|c|c|c|c|c|c|c|c|c|}
\hline
\makecell{\textbf{Dataset}} & 
\makecell{\textbf{Cut-off}\\\textbf{Date}} & 
\makecell{\textbf{Sample Size}\\\textbf{Secure/Vulnerable}} & 
\makecell{\textbf{Projects}} & 
\makecell{\textbf{Data}\\\textbf{Quality}} & 
\makecell{\textbf{Balanced}} & 
\makecell{\textbf{Unique}\\\textbf{CWEs}} & 
\makecell{\textbf{Mean Duplicate}\\\textbf{\% (CWE)}} & 
\makecell{\textbf{Hard}\\\textbf{Neg.\%}} & 
\makecell{\textbf{Strengths}} & 
\makecell{\textbf{Limitations}} \\
\hline
\textbf{Devign}$^{\ast}$ \cite{7} & 2019 & \makecell{\textbf{25,872}\\ (14,978/10,894)}& FFmpeg, QEMU & \mhigh & \cmark & -- & 0.31 & 86.41 & 
\makecell[l]{• Manually validated \\ by experts} & 
\makecell[l]{• Limited projects (2) \\ • Lacks metadata} \\
\hline
\textbf{BigVul}\cite{2} & 2002--2019 & \makecell{\textbf{188,636} \\ (177,736/10,900)} & 348+ & \medium & \xmark & 91 & 11.54 & 19.45 & 
\makecell[l]{• Large dataset \\ • Real-world\\  CVE-linked commits} & 
\makecell[l]{• Heavy label noise \\($\sim$25\% valid) \\ • 12.7\% duplicates \cite{33}} \\ \cdashline{1-9}
BigVul$^{\dagger}$& & \makecell{\textbf{152,256}\\(143571/8685)} &  &  &  & 88 & 1.47 & 19.13 &  &  \\
\hline
\textbf{\textsc{Re}\textlarger{V}\textsc{eal}}$^{\ast}$\cite{9} & 2020 & \makecell{\textbf{22,734} \\ (20,494/2,240)} & Chromium, Debian & \medium & \xmark & -- & 1.59 & 36.13 & 
\makecell[l]{• Real CVEs \\ • Manually validated} & 
\makecell[l]{• Severe class imbalance \\ ($\sim$10:1) \\ • Limited projects} \\
\hline
\textbf{VDISC}$^{\ast}$\cite{4} & 2020 & \makecell{\textbf{1.27M}\\ Vulnerable only} & Juliet, Debian, GitHub etc. & \low & \xmark & 4 & 30.83 & 0.00 & 
\makecell[l]{• Large dataset } & 
\makecell[l]{• Labels via static analyzers \\ • Noisy} \\ 
\hline
\textbf{D2A}$^{\ast}$\cite{40} & 2021 &  \makecell{\textbf{1.30M} \\ (1,276,970/18,653)} & GitHub (6 repos) & \low & \xmark & -- & 84.48 & 6.21 & 
\makecell[l]{• Differential static analysis \\ reduces false positives} & 
\makecell[l]{• Mislabeling \\ • Highly imbalanced} \\
\hline
\textbf{CVEfixes}\cite{36} & 2002--Jul 2024 & \makecell{\textbf{168,089} \\ (159,157/8,932)} & 564 & \mhigh & \xmark & 180 & 30.57 & 9.36 & 
\makecell[l]{• High quality \\ CVE-based labels \\ • Rich metadata} & 
\makecell[l]{• Biased to known CVEs \\ • Not balanced} \\ \cdashline{1-9}
CVEfixes$^{\dagger}$ &  & \makecell{\textbf{139,400} \\ (134,100/5,300)} & &  &  & 180 & 8.07 & 67.5 &  &  \\
% 71.01
% 3.48
\hline
\textbf{CrossVul}\cite{39} & 2021 &  \makecell{\textbf{134,126} \\ (127,242/6,884)} & 498 & \medium & \xmark & 94 & 37.19 & 79.76 & 
\makecell[l]{• Multi-language (40+) \\ • Paired files} & 
\makecell[l]{• Label noise \\ (manual $\sim$48\%) \\ • File-level only} \\ \cdashline{1-9}
CrossVul$^{\dagger}$ &  &  \makecell{\textbf{102,386} \\ (992,202/3,166)} &  &  & & 91 & 10.50 & 79.00 &  &  \\ 
% 33.91
% 13.08
\hline
\textbf{DiverseVul}$^{\ast}$\cite{38} & 2023 & \makecell{\textbf{348,987} \\ (330,492/18,495)} & 797 & \medium & \xmark & 150 & 0.95 & 25.22 & 
\makecell[l]{• Covers 150 CWEs \\ • Deduplicated} & 
\makecell[l]{• Label noise persists \\($\sim$60\% acc.) \\ • Leakage risk} \\
\hline
\textbf{PrimeVul}$^{\ast}$\cite{33} & 2024 & \makecell{\textbf{235,768} \\ (228,800/6,968)}  & libpng, openssl etc. & \mhigh & \cmark & -- & -- & -- & 
\makecell[l]{• Corrected labels \\ • Chronological split} & 
\makecell[l]{• Excludes multi-function/ \\ non-NVD cases} \\
\hline
\textbf{MegaVul}\cite{37} & 2006--Oct 2023 & \makecell{\textbf{353,873} \\ (335,898/17,975)}  & 992 & \mhigh & \xmark & 169 & 2.26 & 0.00 & 
\makecell[l]{• 169 CWEs \\ • Multi-representation \\  (AST, IR, PDG)} & 
\makecell[l]{• Imbalanced \\ • Inherits commit-heuristic \\ limitations} \\ \cdashline{1-9}
MegaVul$^{\dagger}$&  & \makecell{\textbf{309,988} \\ (300,788/9,200)}  &  & & & 168 & 1.10 & 10.09 &  &  \\
\hline
% 37.80
\textbf{VulGate (ours)} & May 2025 & \textbf{\makecell{\textbf{236,663}\\ (119,231/117,432)}} & \textbf{792} & \textbf{\high} & \textbf{\cmark} & \textbf{180} & \textbf{0.44} & \textbf{$\sim$89.0} & 
\makecell[l]{• Cleaned \\ • Structured \\ • Up-to-date \\ • Supports multiple tasks} & 
\makecell[l]{• CWE bias} \\
\hline
\end{tabular}}
\caption{Comparison of Major C/C++ Vulnerability Datasets.}
\label{tab:dataset_unified}

\captionsetup{font=small}
\begin{minipage}{\textwidth}
\scriptsize
\begin{itemize} 
    \item \textdagger denotes the refined/updated version of the original dataset which we incorporate in our dataset VulGate. 
    \item $\ast$ denotes datasets whose commit information is not available for updating.
    \item \textbf{Balance column}: \cmark~ corresponds to vulnerable and secure samples equally represented and \xmark~ corresponds to imbalanced dataset.
    \item \textbf{Data Quality column} (star-based scale):item \high~(five stars): corresponds to large datasets with expert/automated validation and minimal noise; \mhigh~(four stars) indicates large datasets expert/semi-validated labels; \medium~(three stars) reflects datasets with noticeable noise but still widely used and \low~(two star) denotes datasets where labels are automatically generated by the static analyzer 
        \end{itemize}
  \end{minipage}
\end{table*}

% Multiple datasets have been introduced to support the development of ML and DL based vulnerability detection. Devign dataset [38] and REVEAL dataset [18] is a widely adopted benchmark dataset of two projects each, known for its manual verification. However, it lacks fine-grained vulnerability localization and cwe classification etc. Devign is balanced but REVEAL has duplicates and a class imbalance, approx. 10:1.

The construction of specialized datasets for vulnerability detection has progressed in stages, building up from small, manually curated corpora to large-scale, automatically mined datasets. This evolution reflects the research community’s efforts to cater to essential properties including scale, label quality, and diversity Tab.~\ref{tab:dataset_unified}.

% Table~\ref{tab:dataset_unified} summarizes key datasets across multiple dimensions, including size, project coverage, and label quality etc. 

Early benchmarks such as \textbf{Devign} \cite{7} and \textbf{\textsc{Re}\textlarger{V}\textsc{eal}} \cite{9} established the importance of manually validated data with relatively high label quality. However, both these datasets are limited in scope, covering only two projects each, and lacking support for tasks such as CWE classification or fine-grained localization. Moreover, \textsc{Re}\textlarger{V}\textsc{eal} suffers from severe class imbalance\cite{55}, nearly 10:1 ratio of secure to vulnerable classes, making it difficult for ML models to learn robust representations.

% The \textbf{BigVul} dataset \cite{2} is large, real-world C/C++ benchmark up till 2019, it suffers from substantial label noise (manual checks report ~25\% correct and 12.7\% duplicate samples \cite{33}).
To expand in scale, the community shifted toward automated data mining. \textbf{BigVul} is a large, real-world C/C++ dataset comprising linked CVE reports with commits across 348 projects and nearly 190,000 samples \cite{2} . This dataset significantly broadens coverage and is widely used as a benchmark in the community. However, it has shortcomings: one study reveals substantial label noise caused by reliance on commit snapshots \cite{33}, where entire commits are marked as either vulnerable or secure. Manual inspection reports approximately 25\% correct labels, with label duplication exceeding 12.7\%.

% The \textbf{D2A dataset} \cite{40} performs differential analysis across six repositories using Infer \cite{42}. If an alert exists pre‑commit but not post‑commit, the commit is labeled as fixing a vulnerability and labeling the remainder as secure, an assumption that can introduce label noise.
% \textbf{VDISC} \cite{4} very large function-level corpus labeled by static analyzer (Clang/Cppcheck/Flawfinder). Though rich in metadata, its reliance on static analyzer and automated labeling introduces considerable noise.
Datasets \textbf{VDISC} \cite{4} and  \textbf{D2A} \cite{40} rely on traditional vulnerability detection approaches for labeling. VDISC collects over a million vulnerable functions from Juliet, Debian, and GitHub, labeled by static analyzers (Clang/Cppcheck/Flawfinder). Though impressive in scale, the reliance on static analysis inevitably resulted in significant noise. Le et al. reported 50\% of samples are incorrect, and it can adversely affect model evaluation \cite{56}. D2A performs infer-based differential analysis (an optimized ``fast feedback'' form of static code analysis) across six GitHub repositories. A commit is flagged as vulnerable if static analysis warnings resolve when a patch is applied. This approach reduces the false positive rate but may introduce other mislabeling biases due to its reliance on static analysis.

% \textbf{CVEfixes} \cite{36} focuses on real-world CVE and data till June July23,2024 is publishes by CVEfixes, contains moderate label quality labels.
% \textbf{CrossVul} \cite{39} is multi-language vulnerability dataset (40+ languages). Despite being multi-language dataset have label noise(manual accuracy ~48\% \cite{33}).
To address these limitations and improve precision, researchers focused on ground truth: \textbf{CVEfixes} provides high-quality CVE-based labels with rich metadata across 564 projects \cite{36}. However, this dataset is biased towards known CVEs because it only includes publicly disclosed and assigned CVE. Hence, inherently biased toward known and reported cases. \textbf{CrossVul} is a multi-language vulnerability dataset spanning 40+ languages with paired files \cite{39}. However, label reliability is limited: manual scrutiny discovered label accuracy to be 48\% \cite{57}.

% \textbf{DiverseVul} dataset \cite{38} samples spans 150 CWEs and incorporate deduplication and previous datasets BigVul, Devign, \textsc{Re}\textlarger{V}\textsc{eal}, CrossVul, and CVEfixes. It has a label accuracy (manual ~60\% accurate).

% % The DiverseVul paper acknowledges limitations such as label noise due to data duplication, mislabeled commits with whitespace-only, and potential leakage of test set from pretraining data, which could affect model evaluation accuracy.
% The recently introduced \textbf{MegaVul} dataset \cite{37} (2006–Oct2023) has vulnerabilities spanning 169 CWE types from 992 repositories. Its support for multiple code representations, including abstract syntax trees  and intermediate representations. It has rich metadata but still inherits commit‑heuristic assumptions.

% \textbf{PrimeVul} \cite{33} offers corrected from BigVul \cite{2}, CrossVul \cite{39}, CVEfixes \cite{36}, and DiverseVul \cite{38}.It offers high precision, but is biased due to its strict rules exclude multi-function vulnerabilities and non-NVD cases, introducing selection bias and limiting generalization.

\label{sec:literature_review}
We next describe key datasets used for vulnerability detection in the research literature.
% We next describe key datasets and deep learning models used for vulnerability detection in the research literature.

% for DL based vulnerability detection and Model Seclection for vulnerability detection.
% Vulnerability Datasets Related Work
% \subsection{Datasets}

%picture of framework
\begin{figure*}[!ht]
    \centering
    \captionsetup{font=small}
    \includegraphics[width=3.8\textwidth, height=0.38\textheight, keepaspectratio]{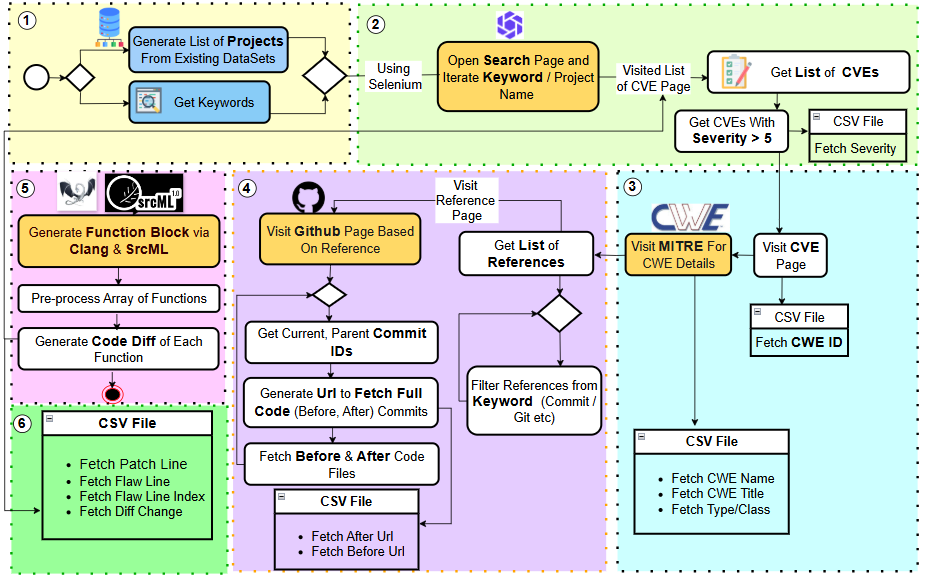}
    \caption{Automated End-to End data collection pipeline for VulGate. The workflow integrates CVE/CWE records, GitHub commits, and function-level parsing to build a structured vulnerability dataset.}
    \label{fig:data_gathering}
\end{figure*}
Attempts were also made to integrate and unify prior datasets. \textbf{DiverseVul} consolidates samples from previous datasets (Devign, BigVul, \textsc{Re}\textlarger{V}\textsc{eal}, CrossVul, and CVEfixes), spanning 150 CWEs with deduplication procedures \cite{38}. This is a diverse dataset, however, the authors report label accuracy of 60\%. 

\textbf{MegaVul} further expands in terms of scale, comprising 17,380 vulnerabilities collected from 992 open-source repositories, and spanning 169 different vulnerability types \cite{37}. This dataset includes multiple code representations (AST, IR, PDG, etc.). However, MegaVul also inherits the limitations of commit-heuristic labeling and is imbalanced. A research study reports 16.7\% of CVEs correspond to ``undecidable'' patches that don’t map cleanly to a single function \cite{54}.

\textbf{PrimeVul}, comprising 6,968 vulnerable and 228,800 secure functions, attempts to address these issues by correcting labels,  carefully filtering prior corpora, and ensuring chronological splits \cite{33}. This yields higher precision but excludes vulnerabilities that are multi-function or not belonging to the National Vulnerabilities Database (NVD), thereby introducing selection bias and limiting generalization. The authors also acknowledge that while the dataset is realistic, it is significantly imbalanced.

The evolution of datasets thus far reflects a trade-off between scale and reliability. Despite significant progress to date, persisting challenges include label noise, duplication, data leakage, lack of diversity, class imbalance, and limited temporal coverage. To address these challenges, we introduce VulGate, our state of the art dataset for vulnerability detection research.

%% file: 3-DataandModel.tex
\section{VulGate: Pipeline and Dataset}
\label{sec:data}
% define centered m-column
\newcolumntype{C}[1]{>{\centering\arraybackslash}m{#1}}

\lstdefinestyle{nobox}{
  basicstyle=\ttfamily\footnotesize,
  breaklines=true,
  columns=fullflexible,
  keepspaces=true,
  showstringspaces=false,
  frame=none,
  backgroundcolor=\color{gray!5},
  escapeinside={(*@}{@*)} % allow LaTeX inside
}

\begin{table*}[ht]
\centering
\begin{adjustbox}{max width=\textwidth}
\begin{tabular}{|C{3.7cm}|C{3.7cm}|C{1.2cm}|C{1.4cm}|C{2.2cm}|C{1.2cm}|C{1.2cm}|C{1.7cm}|C{0.8cm}|C{1cm}|C{2.5cm}|}
\hline

\makecell{\textbf{Vulnerable Code }\\ \textbf{(\textit{Function Before})}} &
\makecell{ \textbf{Secure Code}\\ \textbf{(\textit{Function After})}} &
\makecell{\textbf{CWE}\\ \textbf{ ID}} & \makecell{\textbf{CWE}\\ \textbf{Type}} & \makecell{\textbf{CWE}\\ \textbf{ Description}} &
\textbf{Vulnerable Index} & \textbf{Vulnerable Line No.} & \textbf{Vulnerable Code} &
\textbf{Patch Index} & \textbf{Patch Line No.} & \makecell{\textbf{Patch }\\ \textbf{Code} }\\
\hline

\begin{lstlisting}[style=nobox]
void copy(char *src) {
    char buf[10];
    (*@\textcolor{red}{strcpy(buf, src);} @*)
}
\end{lstlisting} &

\begin{lstlisting}[style=nobox]
void copy(char *src) {
    char buf[10];
    (*@\textcolor{green!60!black}{strncpy(buf, src, sizeof(buf) - 1);} @*)
    (*@\textcolor{green!60!black}{buf[9] = '\textbackslash0';}@*)
}
\end{lstlisting} &

CWE-120 & Buffer Overflow & Buffer Copy without Checking Size & 2 & 3 &
\begin{lstlisting}[style=nobox]
strcpy(buf, src);
\end{lstlisting} & 2 & 3 &
\begin{lstlisting}[style=nobox]
strncpy(buf, src, sizeof(buf) - 1);
buf[9] = '\0';
\end{lstlisting} \\
\hline

\begin{lstlisting}[style=nobox]
void authenticate(char *input) {
    if((*@\textcolor{red}{strcmp(input, "admin")}@*)) {
        printf("Access granted\n");
    }
}
\end{lstlisting} &

\begin{lstlisting}[style=nobox]
void authenticate(char *input) {
    if((*@\textcolor{green!60!black}{strcmp(input, "admin") == 0}@*)) {
        printf("Access granted\n");
    }
}
\end{lstlisting} &

CWE-253 & Logic Error in Auth & Incorrect Check of Return Value & 1 & 2 &
\begin{lstlisting}[style=nobox]
if(strcmp(input, "admin"))
\end{lstlisting} & 1 & 2 &
\begin{lstlisting}[style=nobox]
if(strcmp(input, "admin") == 0)
\end{lstlisting} \\
\hline

\end{tabular}
\end{adjustbox}
% \captionsetup[figure]{font=footnotesize}
% \captionsetup[table]{name=Tab} {font=small}

% \captionsetup[table]{name=Tab, labelfont=normalfont} 
\caption{Illustrative examples of VulGate entries, mapping vulnerable code (\textit{Function Before}) to secure patches (\textit{Function After}) with CWE annotations and metadata.}
\label{table:data}
\end{table*}

% summary of VulGate - the hard numbers
\textbf{VulGate} significantly extends prior datasets in the research literature and includes considerable new content. VulGate consists of 1.36 GB of structured data, comprising 236,663 function-level samples, including 117,432 secure code samples and 119,231 vulnerable code samples. This dataset includes real-world vulnerabilities curated from GitHub commits linked to CVEs, along with CWE mapping and metadata. VulGate also expands temporal coverage of prior datasets from May, 2002 to May, 2025. 

% what makes VulGate unique
Overall, the VulGate dataset spans 792 projects, 7,600 commits, and covers 180 CWE classes, which include memory corruption, input validation, logic errors, and access control. Each code sample includes function-level code from before and after patching, exact line information for vulnerabilities and patches, CWE information, and commit metadata. This fine-grained annotation makes VulGate not only suitable for \textbf{binary classification} but also effective for a range of related applications, including \textbf{vulnerability localization, type classification, and automatic patching}. We believe these unique properties make VulGate one of the most up to date, accurate, and versatile datasets currently available for vulnerability research.

% Table~\ref{table:Top-CWEs} lists the 10 most frequent and critical CWEs in the dataset related to input validation, memory corruption, information disclosure, access control, and memory and resource management. 

\subsection{The VulGate Data Collection Pipeline}
% how to build
We first describe our data collection process. We develop a fully automated data collection pipeline to scrape C/C++ vulnerabilities, integrating multiple sources, including Common Vulnerabilities and Exposures (CVE) records, Common Weakness Enumeration (CWE) taxonomy, and GitHub repositories, and extracting function-level code snippets. The step-by-step process is depicted in Fig.~\ref{fig:data_gathering}.

\newcommand{\encircle}[1]{%
  \tikz[baseline=(X.base)] 
    \node (X) [draw, shape=circle, inner sep=0] {\strut #1};}

The pipeline begins with \encircle{1} selecting C/C++ projects from the aforementioned vulnerability datasets, as well as new projects and extracting relevant \textbf{keywords}, including project name or vulnerability-related terms. Then, \encircle{2} using Selenium based web automation, the system queries public CVE databases with these keywords to retrieve vulnerability reports. To ensure inclusion of high-impact cases, we select CVEs with a severity score greater than 5. 

Next, \encircle{3} we extract the corresponding CWE identifiers for each CVE and query the MITRE CWE repository to enrich the dataset with descriptive metadata, including the CWE title, type, and classification. In parallel, \encircle{4} we parse reference links from CVE entries to locate corresponding GitHub repositories or commits and retrieve vulnerability fixes. We collect both current and parent commit identifiers, thereby enabling access to complete \textit{before} and \textit{after} code versions. 

Then \encircle{5} we process these code snapshots using Clang and srcML to extract structured \textbf{function-level code blocks}. This enables us to precisely analyze code changes, including pinpointing flaw lines, patch lines, and their respective positions, as well as the exact code differences introduced in the fix. Finally, \encircle{6} all collected data, including CVE severity, commit metadata, and detailed code diffs, is recorded in structured CSV files, creating a high-quality dataset for downstream vulnerability research.

Tab.~\ref{table:data} presents VulGate samples, where \textit{Function Before} shows the vulnerable code and \textit{Function After} its secure patch. 
The first case demonstrates a buffer overflow (CWE-120) fixed by replacing \texttt{strcpy} with a bounds-checked \texttt{strncpy}, while the second shows an authentication error (CWE-253) corrected by an explicit comparison in \texttt{strcmp}. 

It took approximately 3 months to collect and pre-process data for VulGate, followed by an additional 100 person-hours for manual verification of subsets by 3 security experts.

\subsection{Introducing Hard Negatives}
% hard negatives

A key contribution of our paper is our incorporation of hard negative samples in the training data. This is an underexplored topic in the literature. By exposing the model to hard negative samples that are semantically similar to vulnerable instances but have different labels, we encourage the model to learn fine-grained distinctions between vulnerable and non-vulnerable samples that are difficult even for human reviewers. This approach significantly reduces false positives and improves cross-domain generalization.

These findings are consistent with work in related domains: for instance, a hard negative sampling approach demonstrated significant improvement in the capacity of models to discriminate and understand code \cite{58}. Likewise, contrastive learning with hard negative samples was found to encourage clear decision boundaries and thereby improve representation quality \cite{59}. Similarly, the effectiveness of retrieval tasks increased when hard negatives were paired with augmentation \cite{60}. 

\subsection{Incorporating Prior Datasets}

% similarity with previous datasets
Tab.~\ref{tab:dataset_unified} summarizes the features of widely used vulnerability datasets, including size, duplication, CWE diversity, CWE distribution, and the percentage of hard negatives in each dataset. 

We make every effort to retain and enhance fundamental features of prior datasets, to ensure that VulGate is compatible with existing research pipelines and benchmark datasets. Like Devign, \textsc{Re}\textlarger{V}\textsc{eal}, and PrimeVul, VulGate provides function-level samples with explicit vulnerable and secure samples. As in the case of BigVul, CVEfixes, and MegaVul, VulGate links CVE and CWE metadata, where each function is linked to real-world vulnerabilities. In line with DiverseVul, VulGate also integrates information from multiple datasets across a diverse set of projects. 

% improvements
We also address key limitations of prior datasets. To rectify issues of duplication, label noise, and poor accuracy encountered in BigVul, CrossVul, and DiverseVul, we apply a strict duplicate removal and cleaning policy. To mitigate class imbalance, encountered in \textsc{Re}\textlarger{V}\textsc{eal} and CVEfixes, VulGate maintains a balanced distribution of vulnerable and patched samples. We also address data leakage concerns other researchers have observed in Devign, BigVul, and DiverseVul \cite{33}, by ensuring clean splits in train/validation/test data free from code overlap.

% [DETAIL MANPOWER AND TIME INVESTMENT IN THIS DATASET?]

Hard negatives vary considerably across the datasets. For instance, Devign 86.41\% shows a very high ratio, but its small scale and restriction to only two projects limit its generalizability. In contrast, large datasets such as MegaVul and VDISC contain virtually no hard negatives, making them less suitable for training. A high proportion 79.76\% is reported in CrossVul, though this figure reflects hard negatives of more than 40 languages and is not directly representative for C/C++ datasets. Notably, VulGate is estimated to contain nearly 89\% hard negatives, combining scale with challenging samples and thus offers a strong foundation for generalization in vulnerability detection. 

Furthermore, unlike prior datasets, VulGate includes rich metadata (CWE ID, CWE description, flaw and patch lines, commit information, and project information). This not only supports multi-task learning for vulnerability detection, but significantly expands the scope and application of this dataset to vulnerability localization, classification, and patch generation.

%% file: 4-Exp.tex
\section{Experimental Evaluation} 

% [Insert short summary of ALL our experiments here - first, we evaluate the effect of increasing context window on a benchmark dataset, which results in a significant performance increase. Next, we bring in vulGate, etc....]

We now describe our models and experiments: first, in Sec.~\ref{sec:model-selection}, we review prior models used for vulnerability detection and rank their performance against a standard dataset (BigVul) used in the literature. In Sec.~\ref{sec:ed}(\textit{Experiment \#1}), we introduce state of the art encoder-decoder models for vulnerability detection and evaluate their performance. These models are fine-tuned for vulnerability detection and demonstrate improved performance compared to prior models.

In Sec.~\ref{sec:context window} (\textit{Experiment \#2}) we investigate the effect of increasing context window size on vulnerability detection. Our results indicate that increasing context window size together with the right architectural choice (encoder-only models) again significantly improves performance.

In Sec.~\ref{sec:evaluation_vulgate} (\textit{Experiment \#3}), we train our new models on our VulGate dataset and showcase more performance gains across diverse projects. In Sec.~\ref{sec:evaluation_crossdata} (\textit{Experiment \#4}, we specifically test generalization capability. We introduce a test set VulGate${^+}$, containing diverse real-world code and synthetic code snippets. Our positive results highlight the importance of hard negatives and balanced training data.

All VulGate datasets and scripts are available at:
\href{https://github.com/Rijha/Gen-Vulgate.git}{Gen-VulGate}.

%followed by a series of experimental evaluations to study the effectiveness of multiple vulnerability detection models. We first evaluate the impact of context window and encoder-only/decoder-only on benchmark dataset, showing that large context window combined with code pre-trained encoder-only leads to significant performance improvement. Next, we introduce our VulGate dataset and demonstrate how it enables generalization and robust performance across diverse projects. Finally, we introduce VulGate${^+}$, containing unseen real-world and synthetic code snippets to further test generalization capabilities. The results highlight the importance of hard negatives and balanced training data.

%Our experiments span multiple datasets (BigVul, VulGate, VulGate$^{+}$), absolute performance numbers are not directly comparable across tables. Instead, we interpret results within each dataset and highlight where differences reflect dataset effects (e.g., Infer’s recall difference between BigVul and VulGate). Also, due to limited computational resources (NVIDIA RTX 2080 GPU), it was infeasible to reproduce all prior models on VulGate. Instead, we evaluated VulGate against representative and widely used baselines such as CodeBERT previous benchmark, and static analyzers as traditional tools. 

\subsection{Review and Evaluation of Detection Models}
\label{sec:model-selection}

We now review the models commonly used for vulnerability detection research. We cover a comprehensive set of approaches ranging from static analyzers and early ML baselines to state of the art DL and LLM based models. For purposes of comparison, in Tab.~\ref{table:benchmark-results}, we enumerate the performance of these models on a common dataset: to maintain consistency, we opt for the BigVul dataset, which is the de facto benchmark in the research literature. We report performance using standard classification metrics: F1-score, Precision (P), and Recall (R). Results for some models (Devign, \textsc{Re}\textlarger{V}\textsc{eal}, and CodeBERT) were replicated and cross-checked against their original publications and found to be in agreement. 

We include static analyzers \textbf{Cppcheck} and \textbf{Infer} in our evaluation, since they represent industry standard baselines. These results also  demonstrate the practical gap between traditional rule-based vulnerability detection and ML/DL approaches, particularly in terms of recall and robustness. Tab.~\ref{table:benchmark-results} shows that Cppcheck achieves 12\% F1, indicating that it misses the vast majority of vulnerable samples. Infer performs slightly better with 19.5\% F1 but still exhibits very low overall effectiveness, showing the limitations of static analyzers.
\begin{table}[t]
\centering
\begin{tabular}{|c|c|p{0.648cm}|p{0.648cm}|p{0.648cm}|}
% \begin{tabular}{|p{1.65cm}|p{2.3cm}|p{0.68cm}|p{0.68cm}|p{0.68cm}|}
\hline
\textbf{Category} & \textbf{Techniques/Models} & \textbf{F1} & \textbf{P} & \textbf{R} \\
% Infer \cite{42} and Cppcheck \cite{21}
\hline
\multirow{2}{*}{\makecell{\textbf{Static}\\\textbf{Analyzer}}} & Cppcheck \cite{21} & 12\% & 10\% & 15\% \\
\cline{2-5}
                                & Infer \cite{42} & 19.5\% & 15\% & 28\% \\
\hline
\multirow{2}{*}{\textbf{ML}}& BoW+RF\cite{5,6} & 25\% & 48\% & 17\% \\
\cline{2-5}
                                & Russell et al.\cite{4} & 24\% & 16\% & 48\% \\
\hline
\multirow{2}{*}{\textbf{DL}} & VulDeePecker\cite{3} & 19\% & 12\% & 49\% \\
\cline{2-5}
                                & SySeVR\cite{8} & 27\% & 15\% & 74\% \\
\hline
\multirow{2}{*}{\textbf{GNN}}& Devign\cite{7} & 26\% & 18\% & 52\% \\             
\cline{2-5}
                                & \textsc{Re}\textlarger{V}\textsc{eal}\cite{9} & 30\% & 19\% & 74\% \\
\cline{2-5}
                                & IVDetect\cite{10} & 35\% & 23\% & 72\% \\ 
\hline 
\multirow{2}{*}{\textbf{Decoder}}& {\makecell{\textbf{CodeGPT-2} \\(context window:1024)} }& 90.45\% & 97\% & 84.45\% \\
\cline{2-5}
                                & {\makecell{\textbf{CodeLlama}\\(context window:1024)}} & 79.73\% & 79.34\% & 80.13\% \\                                 
\hline
\multirow{2}{*}{\textbf{Encoder}}& CodeBERT\cite{1} & 91\% & 97\% & 86\% \\      
\cline{2-5}
                                & {\makecell{\textbf{UniXcoder-Base}\\ (context window:512)}} & 84.68\% & 84.32\%& 85.04\% \\
\cline{2-5}
                                & {\makecell{\textbf{UniXcoder-Base}\\ (context window:1024)}} & 94.23\% & \textbf{97.28\%} & 91.37\% \\
\cline{2-5}
                                & {\makecell{\textbf{UniXcoder-Base-Nine}\\ (context window:512)}} & 88.8\% & 87.9\% & 89.7\%\\
\cline{2-5}
                                & {\makecell{\textbf{UniXcoder-Base-Nine}\\ (context window:1024)}} & \textbf{94.73\%} & 96.74\% & \textbf{92.8\%} \\
\hline

\end{tabular}
\medskip

\caption{F1-score, Precision (P), and Recall (R) on the BigVul benchmark dataset. \textbf{Bold values} denote our experimental runs, with the best results in each column also highlighted in bold. Results of BoW+RF, Russell, VulDeePecker, SySeVR and IVDetect are taken from \cite{1}}
\vspace{-0.5 cm}
\label{table:benchmark-results}
\end{table}

Early ML approaches relied primarily on shallow features, such as bag-of-words (BoW) and n-gram token frequency, often coupled with classification algorithms like random forests (RF) \cite{5,6}. These solutions are easy to deploy and computationally efficient but, as Risse~\textit{et al}. report, this reliance on lexical features results in failure to reason about semantic behavior of code and achieves limited recall \cite{50}. Consistent with this, the BoW+RF baseline achieves a relatively high precision of 48\% but very poor recall of 17\%, highlighting that such models tend to overfit due to reliance on superficial features.

A pivotal shift in vulnerability detection occurred with use of learning-based techniques, first developed by \textbf{Russell} et al. \cite{4}, who used a deep feature representation learning technique that directly interprets lexed source code. However, their use of synthetic code fragments, i.e., ``good'' and ``bad'' words injected into source code samples limits real-world applicability and also significantly impacts the ability of the model to generalize. The result show F1 of 24\% with recall 48\%, demonstrating improved performance but still weak overall performance.

This was followed by the first wave of DL solutions: \textbf{VulDeePecker} used bidirectional LSTMs to model code sequences for vulnerability detection \cite{3}. The model achieves only 19\% F1, reflecting many false positives and false negatives. \textbf{SySeVR} integrated enriched vulnerability representation by combining syntax (\textit{Sy}) and semantics (\textit{Se}) slicing \cite{8}. These models outperformed previous approaches with recall of 74\% but precision remained low at 15\%, indicating that models struggled with complex long-range dependencies in code.

\textbf{Devign} \cite{7}, a GNN-based model achieves 26\% F1 but still suffers from imbalanced precision and recall. \textbf{\textsc{Re}\textlarger{V}\textsc{eal} }\cite{9}, a GNN based model continues this trend, achieving 30\% F1 with high recall 74\% but low precision 19\%. Finally, \textbf{IVDetect} captures long-term dependencies \cite{10}, reaching 35\% F1, the best in this group, but still far below LLM-based approaches.

The results in Tab.\ref{table:benchmark-results} highlight a clear progression. Vulnerability detection solutions in prior work, such as VulDeePecker, SySeVR, and IVDetect show limited performance (achieving F1 scores in the range of 19–35\%), largely due to their reliance on handcrafted program representations and limited ability to capture semantic context.

\subsection{Experiment \#1: Encoder vs Decoder Architectures}
\label{sec:ed}

% Intro + models tested

% Implementation details

% Decoder family (intro → results → discussion)

% Encoder family (intro → results → discussion)

% Overall conclusion + segue to next experiment

Pre-trained transformer architectures mark another breakthrough in vulnerability detection by bringing advances from NLP tasks into the domain of code understanding and software security. To systematically investigate the role of architectural choices, we undertake an evaluation exercise using decoder-only and encoder-only families, specifically \textbf{CodeGPT-2}, \textbf{CodeLlama}, \textbf{CodeBERT}, \textbf{UniXcoder-Base}, and \textbf{UniXcoder-Base-Nine}. The results are listed in Tab.\ref{table:benchmark-results}.

\textbf{Implementation details and experimental setup:} All models were implemented in PyTorch \cite{62} with the HuggingFace Transformers \cite{63} and DeepSpeed library \cite{64}. For the encoder models (CodeBERT, UniXcoder-Base and UniXcoder-Base-Nine),  we use the standard configuration i.e., 12 encoder blocks, a hidden size of 768, and 12 attention heads. Following common practice \cite{34}, we fine-tune using the AdamW optimizer with a learning rate of $2\times10^{-5}$, applying a linear decay schedule. Decoder-only models CodeGPT-2 and CodeLlama are finetuned using deepspeed and parameter-efficient fine-tuning (LoRA) respectively. A learning rate of \(1\times10^{-4}\), warmup steps, gradient accumulation, mixed precision (fp16/bf16), and quantization (4-bit) where supported. 

For all models, training was run for up to 15 epochs. We used a batch size of 30 per GPU and early stopping based on validation F1-score. Cross-entropy loss was used as the training objective. All experiments were conducted on NVIDIA RTX 2080 graphic card. Fine-tuning typically converged within 3–12 epochs.

%All models are benchmarked on BigVul, a large-scale dataset for real-world vulnerability detection in C/C++ projects, using the same training, validation, and test splits as in \cite{1} for fairness.

\textbf{Results:} Decoder-based models such as \textbf{CodeGPT-2} and \textbf{CodeLlama} were included to evaluate whether autoregressive architectures designed for next-token prediction can be effective for vulnerability detection. These models were chosen because they are pre-trained on code and are powerful for generative tasks. Results indicate they underperformed compared to encoder-based models, even with optimization strategies (deepspeed, mixed precision and quantization) and LoRA.

% This performance can be attributed partly to the pretraining objective i.e., causal language modeling encourages autoregressive, next word generation oriented representations that are less directly aligned with sequence-level classification.

Although CodeGPT-2 and CodeLlama are large models having 1.5B and 7B parameters respectively and they benefit from large scale pretraining, the relatively weak performance, especially of CodeLlama, can be attributed to a misalignment between the architecture and our objective, i.e. classification. Decoder-only models are primarily optimized for generative tasks such as code generation rather than discriminative classification. This makes them less effective at capturing bidirectional dependencies and fine-grained structural similarities that are critical for vulnerability detection (this point has been noted in recent surveys on LLMs in software security \cite{66}). Additionally, the scale of CodeLlama compared to the available labeled data further decrease performance, since larger models generally require very large amounts of data to adapt effectively \cite{65}.

% This performance contrast supports the conclusion that encoder-based architectures, with their bidirectional attention and masked language modeling objectives, are inherently better suited for vulnerability detection tasks.

% and \textbf{GraphCodeBERT} \cite{35} 
Encoder-only models such as \textbf{CodeBERT} represent bidirectional models trained on source code \cite{34} and transfer well to downstream vulnerability classification tasks. For instance, CodeBERT achieves 91\% F1 and 86\% recall on BigVul, outperforming all other solutions in the literature. We replicate these results and our findings are consistent with prior reports \cite{1}.

\textbf{UniXcoder}\cite{13} and \textbf{UniXcoder-Base-Nine}\cite{14} consistently outperform other models, likely due to their pretraining strategy which is explicitly tailored for code comprehension. These models were pre-trained on NL-PL pair of CodeSearchNet dataset containing six programming languages, i.e., Java, Ruby, Python, JavaScript, Go. Additionally, UniXcoder-Base-Nine is further trained on 1.5 million NL-PL pairs of C, C++ and C\#. 

The performance improvement can be attributed to encoder masked language modeling (MLM) pretraining and bidirectional attention which allows the model to attend to whole sequences, capturing structural dependencies effectively within source code, making them more suitable for vulnerability detection tasks.

From this experiment we concluded that encoder-only models are generally more effective for software vulnerability detection. To further disentangle the effect of model architecture from input length, we undertook a follow-up experiment varying the context window size.

% To systematically investigate architectural choices, we fine-tuned both encoder and decoder families, specifically \textbf{CodeBERT}, \textbf{UniXcoder-Base}, \textbf{UniXcoder-Base-Nine}, \textbf{CodeGPT-2} and \textbf{CodeLlama}. We benchmarked these models on BigVul, a large scale benchmark for real-world vulnerability detection in C/C++ projects. To ensure fairness, we adopted the same training, validation, and test splits as used in \cite{1}.

% To establish comparability with prior work, we include five baselines (VulDeePecker, Russell, BoW+RF, SySeVR and IVDetect) in our comparison taken from the \cite{1} study. Additional results for static analyzers (CppCheck, Infer) and (Devign, \textsc{Re}\textlarger{V}\textsc{eal}, CodeGPT-2, CodeLlama, CodeBERT, UniXcoder variants) were produced in our experiments Tab.\ref{table:benchmark-results}. 

\subsection{Experiment \#2: Context Window Size}
\label{sec:context window}

% In this section, to address the limitations, we evaluate  multiple encoder and decoder based models for software vulnerability detection. 
% In this section, our goal is to assess their effectiveness across real-world benchmarks and highlight the impact of context length, architecture, and domain-specific pretraining.
\begin{table}[t]
\centering
% \resizebox{\columnwidth}{!}{%
\begingroup
\small 
\begin{tabular}{|c|c|c|c|}
\hline
\textbf{Split} & \textbf{Total Samples} & \textbf{Class 0} & \textbf{Class 1} \\
\hline
Train & 189,330 & 95,295 & 94,035 \\
\hline
Validation & 23,666 & 11,995 & 11,671 \\
\hline
Test & 23,667 & 11,941 & 11,726 \\
\hline
\textbf{Total} & \textbf{236,663} & \textbf{119,231} & \textbf{117,432} \\
\hline
\end{tabular}%
\endgroup
% }
\caption{Distribution of samples in VulGate vulnerability dataset used for fine-tuning. Class 0 represents non-vulnerable samples and Class 1 represents vulnerable samples.}
\label{tab:dataset-splits}
\end{table}
We vary input length from 512 to 1024. CodeBERT maximum context window size is 512, whereas both variants of UniXcoder support larger context window of 1024. We experimented with both variants of UniXcoder with context length 512 and 1024.

\textbf{Implementation details and experimental setup:} To increase context window of decoder-only models on GPU resources (NVIDIA RTX 2080 GPU and Intel Xeon Gold 6230 CPU), we used \textbf{DeepSpeed} Library and optimize training of CodeGPT-2 \cite{11} and applied \textbf{parameter-efficient fine-tuning (PEFT)} via \textbf{LoRA} for CodeLlama\cite{12} fine-tuning. In contrast, UniXcoder variants with 1024-token context window were fully fine-tuned, which allowed us to adapt all model parameters for the vulnerability detection task.

%Analysis of Results
\textbf{Results:} As observed in Tab.~\ref{table:benchmark-results}, for both UniXcoder variants, extending the input length from 512 to 1024 tokens consistently improved performance. Specifically, UniXcoder-Base improved from 84.68\% F1 (context window:512) to 94.23\% F1 (context window:1024). Similarly, UniXcoder-Base-Nine improved from 88.8\% F1 (context window:512) to 94.73\% F1 (context window:1024).

These results confirm that many real-world vulnerabilities span contexts exceeding 512 tokens, and that extending the input window allows the model to capture the broader control and data flow required for correct classification. In contrast, CodeBERT is constrained to 512 tokens, which limits its ability to reason over long functions despite achieving competitive performance (91\% F1).

\subsection{Experiment \#3: Benchmark Evaluation on VulGate}
\label{sec:evaluation_vulgate}

We now investigate the performance of our leading models on the VulGate dataset.

\textbf{Implementation details and experimental setup:} Tab ~\ref{tab:dataset-splits} summarizes the train/validation/test split with an 80:10:10 ratio, ensuring class balance. As we noted earlier, to strengthen robustness and generalizability, VulGate explicitly incorporates hard negative samples.

% the code snippets that are semantically similar but with different labels, which exposed the model to non-vulnerable samples that closely resemble vulnerable ones. 

%The existing study demonstrate remarkable results on training and testing datasets, a fundamental limitation is that this performance does not generalize to unseen codebase \cite{33} due scarce and mislabeled data and relatively shorter context windows that restricted the ability to capture long-range code dependencies. So, we fine-tuned and evaluated our top performing models on \textbf{VulGate}, a refined and balanced benchmark for real-world vulnerability detection. This setup allows us to assess model generalization against both traditional static analyzers and LLMs. 

\begin{table}[t]
\centering
\begin{tabular}{|c|c|c|c|c|}
% \begin{tabular}{|p{1cm}|p{2.5cm}|p{0.68cm}|p{0.68cm}|p{0.68cm}|}
\hline
\textbf{Dataset} & \textbf{Techniques/Models} & \textbf{F1} & \textbf{P} & \textbf{R} \\
\hline
VulGate & Cppcheck & 29.0\% & 54.0\% & 20.0\% \\
\cline{2-5}
            & Infer    & 42.0\% & 26.0\% & 98.0\% \\
\cline{2-5}
            & CodeBERT & 85.9\% & 83.2\% & 89\% \\
\cline{2-5}
            & UniXcoder-Base& 87\% & 85.4\% & 88.7\% \\
\cline{2-5}
            & UniXcoder-Base-Nine & 88.9\% & 87.7\% & 90.0\% \\
 
% VulGate & UniXcoder-Base-Nine & 88.86\% & 87.73\% & 90.02\% \\
  
\hline
\end{tabular}
\medskip
\caption{Results of F1-score, Precision (P), and Recall (R) on the Refined Vulnerability dataset on our best performing model, previous SOTA CodeBERT \cite{1} and Static Analyzer (Infer \cite{42} and Cppcheck \cite{21})}
\label{table:my-results}
\end{table}

 % Our carefully designed fine-tuning strategy incorporates our refined diverse data VulGate, containing 792 projects, correct label, with selected samples (approx. 1000) verified by security experts, hard negatives, and class balance ensured that model performance shows true generalization rather than memorization.

In the training data, we introduce 792 projects containing 189,330 samples with equal number of vulnerable and non-vulnerable samples. The dataset contains samples with more than 90\% cosine similarity, thereby forcing the model to learn the distinction between vulnerable and non-vulnerable patterns. This approach also reduces false positives and ensures that model performance shows true generalization rather than memorization. The selected samples (approx. 1000) verified by security experts.

% Our carefully designed fine-tuning strategy incorporates our refined diverse data VulGate, containing 792 projects, correct label, with selected samples (approx. 1000) verified by security experts, hard negatives, and class balance ensured that model performance shows true generalization rather than memorization.

%Together, these strategies produced model checkpoints that are reliable even in previously unseen settings, as shown in Tab.\ref{table:generalization-results}.

\textbf{Results:} As depicted in Tab.\ref{table:my-results}, the encoder and decoder models significantly outperform static analyzers, Cppcheck and Infer. The UniXcoder family surpassed CodeBERT, which served as a strong baseline in prior work. UniXcoder-Base-Nine achieves an F1 score of 88.9\%, precision of 87.7\%, and recall of 90.0\% compared to CodeBERT’s 85.9\% F1. These results confirm that our fine-tuning strategy captures vulnerability patterns better than both traditional analyzers and pre-trained baselines. 

The performance of VulGate is generally lower than on BigVul. This is because VulGate is intentionally harder and more realistic. It removes duplicates, maintains a balance distribution of vulnerable and non-vulnerable samples, spans diverse projects, includes carefully verified labels and incorporates a large number of hard negatives which forces the model to learn true semantic features rather than superficial patterns. These characteristics make VulGate a stricter yet a reliable benchmark for evaluating vulnerability detection models, while also improving generalizability as shown in Tab.\ref{table:generalization-results}.

We observe a contrast for static analyzers: Infer’s recall rises to 98\% on VulGate as compared to 28\% on BigVul, but precision drops to 26\%. This highlights the impact of VulGate’s balanced and carefully cleaned samples, which increases the likelihood that static analyzers trigger on vulnerable samples. However, this comes at the cost of over-flagging many secure snippets. Thus, high recall here should not be interpreted as improved analyzer performance but rather as sensitivity to dataset characteristics.

\begin{figure}[t]
    \centering
    \captionsetup{font=small}
    \includegraphics[width=\columnwidth, keepaspectratio]{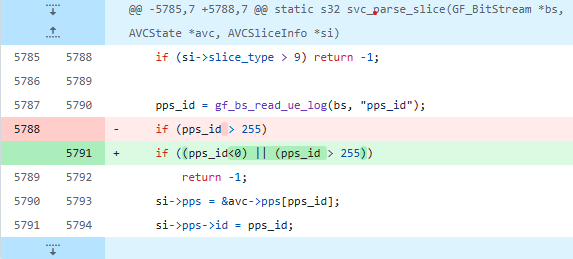}
    \caption{A visual example of vulnerability (CVE-2021-40568) correctly identified by UniXcoder-Base-Nine: Vulnerable lines in \texttt{svc\_parse\_slice}. 
    The check on \texttt{pps\_id} is insufficient, leading to a potential buffer overflow.}
    \label{fig:svc_parse_slice_vuln}
\end{figure}

A visual example of vulnerability instances correctly identified by UniXcoder-Base-Nine but missed by other models (including CodeBERT \cite{1}) is shown in Fig.~\ref{fig:svc_parse_slice_vuln}.
The snippet is from the gpac project (CVE-2021-40568) and corresponds to CWE-120 (Buffer Copy/Access without Checking Size) in \texttt{svc\_parse\_slice}. This vulnerability may be exploited using a crafted MP4 file in GPAC  \(\leq\)  1.0.1 and may result in denial of service, arbitrary code execution, or privilege escalation. The bounds check only ensures \texttt{pps\_id \(\leq\)  255}, but does not account for the actual size of the \texttt{$avc->pps$} array. As a result, the subsequent indexing operation may access memory out-of-bounds and cause a buffer overflow.

\subsection{Experiment \#4: Cross Dataset performance}
\label{sec:evaluation_crossdata}
% \subsection{Generalization to Unseen Datasets}
We now evaluate the generalization capabilities of CodeBERT and UniXcoder-Base-Nine beyond in-distribution data.

\textbf{Implementation details and experimental setup:} We enhance our dataset with a test set specifically crafted for this task and strictly excluded from training, validation, or testing. Denoted as \textbf{VulGate$^{+}$}, this test set comprises:  

%, we compare CodeBERT (previous benchmark) and UniXcoder-Base-Nine on VulGate$^{+}$ datasets. The \textbf{VulGate} benchmark consists of two components: (i) our cleaned refined dataset \textbf{VulGate} containing prior datasets including BigVul, and (ii) three curated datasets strictly excluded from training, validation, or testing denoted as \textbf{VulGate$^{+}$}. The VulGate$^{+}$ collection includes:
\begin{itemize}
    \item A Collection of recent \textbf{Linux vulnerabilities} containing 78 vulnerable and 1,805 non-vulnerable samples from 2024 and 2025,
    \item A synthetic but realistic dataset of 150 vulnerable and 150 non-vulnerable samples generated using \textbf{Claude} \cite{41} and manually verified by three security experts (consisting of approximately 25 hours of review), and
    \item The \textbf{PrimeVul dataset}, which also builds on prior vulnerability datasets but employs a distinct cleaning methodology \cite{33}. For this set, we ensure no overlap with our training, validation, or test sets, resulting in 7,422 vulnerable and 7,164 non-vulnerable unique samples.
\end{itemize} 

\begin{table}[t]
\centering
\caption{F1-score, Precision (P), and Recall (R) of CodeBERT and UniXcoder-Base-Nine on VulGate$^{+}$ datasets.}
\resizebox{\columnwidth}{!}{%
\begin{tabular}{|c|c|c|c|c|}
\hline
\textbf{Dataset} & \textbf{Models} & \textbf{F1} & \textbf{P} & \textbf{R} \\
\hline
\textbf{Linux Data}& CodeBERT \cite{1} & 75.3\% & 69.5\% & 82.1\% \\
                      & UniXcoder-Base-Nine & 76.4\% & 93.7\% & 64.1\% \\
\hline
\textbf{PrimeVul Data} & CodeBERT \cite{1} & 87\% & 87\% & 87\% \\
                       & UniXcoder-Base-Nine &89.14\% & 87.1\% & 91.3\% \\
\hline
\textbf{Claude Data} & CodeBERT \cite{1} & 17.25\% & 53.1\% & 10.3\% \\
                     & UniXcoder-Base-Nine & 64.8\% & 92\% & 50\% \\
\hline
\end{tabular}%
}
\label{table:generalization-results}
\end{table}

\textbf{Results:} Tab.\ref{table:generalization-results} shows that across all test sets UniXcoder-Base-Nine consistently outperforms CodeBERT \cite{1}. On Linux data, where CodeBERT achieved 75.3\% F1 score, UniXcoder-Base-Nine is slightly outperformed with 76.4\%, though recall remained limited due to the small dataset size. On PrimeVul data, our  UniXcoder-Base-Nine achieved 89.1\% F1 score compared to CodeBERT’s 87\% with a higher recall of 91.3\%, showing its ability to capture more true vulnerabilities. Most importantly, on the expert verified Claude dataset which is the toughest due to its synthetic yet realistic nature, where CodeBERT collapsed to just 17.25\% F1 score, whereas UniXcoder-Base-Nine sustained 64.8\% F1 score, representing a 3.7× relative improvement.

%% file: 5-discussion.tex
\section{Further Discussion}

%The results thus far validate our hypotheses: transformer architectures, pre-trained on code, particularly encoder models like UniXCoder-Base-Nine significantly outperform prior models in the literature. Context window size has a significant impact on performance. Furthermore, training models on a clean, diverse, and balanced dataset is critical for generalization capability. 

%\subsection{Generalization}
Prior literature consistently shows that detection models suffer from catastrophic performance degradation when tested on cross-data or unseen projects, rendering them unfit for real-world use.

For instance, Chakraborty et al. demonstrate that on average, the performance of pre-trained models drops by about 73\% when tested on real-world vulnerabilities \cite{9}. For example, VulDeePecker's precision drops from 86.9\% to just 11.1\%. Moreover, even after retraining on real-world data, models perform on average 54\% lower than initially reported. Similarly, Yizheng et al. \cite{38} found that GNNs, encoders, decoders, and transformers achieve 49\% F1 score on seen projects but this score drops drastically to only 9.4\% when tested on unseen projects.

In this sense, our positive scores in Tab.\ref{table:generalization-results} represent a significant breakthrough in generalization. Even our lowest F1 score (64.8\% on Claude data) significantly surpasses the best reported generalization scores in the literature.

% Our generalizability results Tab.\ref{table:generalization-results} may not appear as strong as in-distribution results presented in Tab.\ref{table:benchmark-results} and \ref{table:my-results}, they provide a true depiction of real-world performance under distributional shifts. An important point is to note that our model exhibit only modest declines compared to prior work. 

The appropriate comparison to highlight our results is a recent benchmark study \cite{33}: Yangruibo et.al test CodeBERT and UniXcoder on the BigVul dataset, achieving F1 scores of 62.88\% and 65.46\%, whereas on the (unseen) PrimeVul dataset, these scores drop dramatically to 20.86\% and 21.43\% respectively. The performance loss in both cases exceeds 40\%.
%reported 42.02\% and 44.03\% in both respectively.

In contrast, our pretrained CodeBERT and UniXcoder-Base-Nine models achieve F1 scores of 91\% and 94.7\% on BigVul (Tab.\ref{table:benchmark-results}) and 87\% and 89.1\% respectively on PrimeVul (Tab.\ref{table:generalization-results}). Our performance drop is 4-6\%, an order of magnitude less than results achieved by Yangruibo et.al. and, to the best of our knowledge, the highest generalization results reported in the literature to date.

These results validate our approach: transformer architectures, pre-trained on code, particularly encoder models like UniXCoder-Base-Nine significantly outperform prior models in the literature. Context window size has a significant impact on performance. Furthermore, training models on a clean, diverse, and balanced dataset is critical for generalization capability.

%% file: 6-end.tex
\section{Conclusion and Future Work}

In this study, we undertook a comprehensive vulnerability detection exercise using LLMs. We specifically addressed the critical challenge of poor generalization. Our experiments demonstrate that model architecture, context window size, training data quality, and hard negative mining strategies all significantly impact and improve generalization.

We conducted extensive experiments with encoder-only and decoder-only models. Among them, UniXcoder-Base-Nine consistently outperformed alternatives such as CodeBERT, CodeGPT-2, and CodeLlama across both benchmark and unseen datasets. Moreover, increasing context window size enabled models to identify vulnerabilities distributed over larger functions. Our generalization results represent a dramatic improvement over prior results attained in the literature.

In this paper, we also introduce VulGate, our custom dataset, which addresses the shortcomings and limitations of prior datasets. We present a complete pipeline that removes mislabeled and duplicate samples, incorporates recent vulnerabilities up to May 2025, adds rich metadata, integrates hard negative samples, and includes a dedicated test set for generalization testing. To the best of our knowledge, VulGate is the most advanced dataset for vulnerability detection to date, and is also useful for related applications of code classification, localization, and patching.

In future work, we intend to extend our approach along three directions: first, evaluation across multiple programming languages will verify cross-language generalization. Second, exploration of instruction-finetuning using information in the VulGate metadata to further improve vulnerability detection along with automated remediation. Finally, we intend to investigate hybrid training that incorporates contrastive learning to amplify the benefits of hard negatives and further improve generalization.

We hope our effort contributes to the development of effective and reliable automated vulnerability detection solutions.